\documentclass[twocolumn, a4paper, showpacs]{revtex4}
\usepackage{dcolumn,amsmath,xspace}
\usepackage{graphicx}
\usepackage{bm}
\usepackage{color}

\begin{document}
\title{High magnetic field induced charge density waves and sign reversal of the Hall coefficient in graphite}

\author{Amit Kumar$^1$*, Jean-Marie Poumirol$^1$, Walter Escoffier$^1$*, Michel Goiran$^1$, Bertrand Raquet$^1$, and Jean Claude Pivin$^2$}

\affiliation{$^1$ Laboratoire National des Champs Magnétiques Intenses (LNCMI), Universit\'e de Toulouse, INSA, UPS, CNRS-UPR3228, 143 av. de rangueil, 31400 Toulouse, France}%
\email{amit.kumar@lncmi.cnrs.fr,  walter.escoffier@lncmi.cnrs.fr}
\affiliation{$^2$ Centre de Spectrométrie Nucléaire et de Spectrométrie de Masse (CSNSM), IN2P3-CNRS, Batiment 108, 91405,  Orsay Campus, France}%

\begin{abstract}
We report on the investigation of magnetic field induced charge density wave and Hall coefficient sign reversal in a quasi-two dimensional electronic system of highly oriented pyrolytic graphite under very strong magnetic field. The change of Hall sign coefficient from negative to positive occurs at low temperature and high magnetic field just after the charge density wave transition, suggesting the role of hole-like quasi-particles in this effect. Angular dependent measurements show that the charge density wave transition and Hall sign reversal fields follow the magnetic field component along the c-axis of graphite. 
\end{abstract}


\maketitle

Graphite is a typical layered material made from stacked single atomic layers of carbon atoms, named graphene. A few years ago, it has been possible to extract a graphene layer out of bulk graphite and to individually address its electronic properties \cite {Novoselov04}. The discovery of such an ultimate two-dimensional system has attracted tremendous interest in condensed matter physics, especially due to its distinctive band structure and potential applications in nano-electronic devices. In the meantime, such progresses have also triggered a revival interest in the mother material, graphite, pointing out its quasi-two dimensional character which shares some similarities with graphene and which were skipped in the past. The observation of massless electronic charge carriers (Dirac fermions) \cite{Orlita08,Zhou06,Igor06}, quantum Hall effect \cite{Igor06, Kopelevich03} or even possible fractional quantum Hall effect \cite{Kopelevich09} in Highly Oriented Pyrolytic Graphite (HOPG) are just a few examples suggesting the unexplored physics of this material.

Graphite rapidly reaches the quantum limit at moderate magnetic field ($>$8T) due to the in-plane light effective carrier mass and low carrier density. Beyond this magnetic quantum limit no drastic physical effects are expected to occur. However, earlier magneto-transport studies in graphite have clearly revealed an abrupt increase of the in-plane magneto resistance ($R_{xx}$) when a very high magnetic field ($>25$ T) is applied parallel to the c-axis. The occurrence of this effect in a restricted temperature range ($1K<T<10K$) \cite{Iye82, Iye85,Yaguchi98} has been qualitatively explained in terms of a field-induced electronic phase transition by Yoshioka and Fukuyama \cite{Yoshioka81}, later referred to as the YF model, where a charge density wave (CDW) instability develops along the c-axis, induced by the Landau level formation in the highly anisotropic energy band structure of graphite. The YF model describes well the phase transition boundary in the (B,T) diagram, but fails to account for the magnitude and fine additional features of the magneto-resistance jump at the transition \cite{Iye85,Yaguchi98}. Also, the discussion was enriched in \cite{Yaguchi01} where the out-of-plane magneto-resistance exhibits a much larger change as compared to the in-plane magneto-resistance and suggests the occurrence of a spin density wave along the c-axis. So far, the high-field induced anomaly in magneto-transport measurements is not fully understood and requires further developments.

To date, most of the high magnetic field measurements have been reported on Kish or natural graphite and Hall measurements were not discussed in details. Simultaneous measurements of longitudinal (in-plane) resistivity and Hall resistance is expected to give more insight on the physical phenomena that occur beyond the ultra-quantum limit in graphite. Indeed, the Hall effect is a very sensitive tool for investigating the charge carriers' density, as well as their sign and dynamics. In this work, we report on transport and Hall resistance measurements on HOPG in pulsed magnetic field ($\pm$ 57 T). The occurrence of a field-induced resistance anomaly and the Hall sign reversal are investigated in a wide temperature range (1.6 K - 300 K) and for different orientations of the magnetic field.

The HOPG used in this study is of type ZYB with small mosaic angle ($\approx$ 0.80) and was supplied from SPI, USA \cite {SPI}. The HOPG samples were cut into a suitable size and freshly cleaved before mounting into the experimental setup. Electrical  contacts were prepared with silver epoxy. Two HOPG samples were studied (named sample A and B) with typical lateral dimensions of a few hundred microns. The thickness was roughly estimated to 10 $\mu$m by cross-sectional scanning electron microscopy (SEM) measurement. Assuming a uniform distribution of the injected current through the sample, the typical samples' resistivity is 0.33$\pm$0.2 m$\Omega$cm at room temperature and zero magnetic field. The estimated resistivity is about 10-100 times higher than the best quality HOPG \cite{Kopelevich03, Kopelevich09} and Kish graphite \cite{Iye82}. Generally, a higher resistivity is expected in thin graphite flakes \cite{Barzola08} or graphite with small grain size \cite{Zhang04}. The high resistivity found in the present case may also be due to damages made to the sample during preparation (contamination, inhomogeneous cleavage or damages when making contacts). Magneto-transport experiments were repeated with reversing the magnetic field and averaging the signals asymmetrically in order to cancel the effects of voltage probes misalignment. 

\begin{figure*}[!ht]
\includegraphics[width=.8\linewidth]{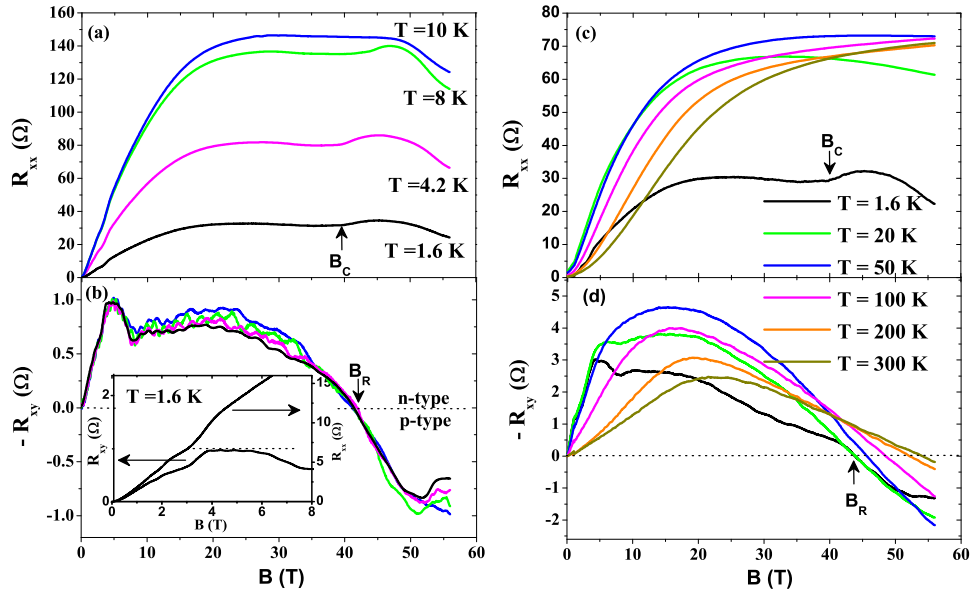}
\caption{\label{fig1} (a) and (c) : longitudinal resistance ($R_{xx}$) and (b) and (d) : Hall resistance ($R_{xy}$) for sample A and B, respectively. Inset figure \ref{fig1}(b) focuses onto the low magnetic field range, where SdH oscillations and onset of Hall plateaus are shown at low temperature.}
\end{figure*}

Figure \ref{fig1} (a, b) shows the in-plane longitudinal resistance ($R_{xx}$) and Hall resistance ($R_{xy}$) at different temperatures comprised between $1.6K$ to $10K$, up to 57 T. We first focus on the low magnetic field regime ($<$9T) where the appearance of plateau-like features and SdH oscillations can be seen in the Hall and longitudinal resistance, respectively (see inset of Fig. \ref{fig1}-b). Both features are consistent with previously reported magneto-transport measurements in high quality HOPG \cite{Kopelevich03,Igor06} and are attributed to the formation of Landau levels. When plotted against $1/B$, SdH oscillations are periodic with $\Delta$($B^{-1}$) $\approx $ 0.2 $T^{-1}$ in agreement with Kopelevich {\it et. al.} \cite{Kopelevich09}, yielding an estimated carrier density of $n\approx 3\times$$10^{11}\ cm^{-2}$. A detailed investigation suggests that such measured carrier concentration is most probably not intrinsic to graphite, but comes from defects in the lattice structure or from impurities \cite{Arndt09} which act as electron/hole donors. Magneto-optical \cite{Orlita08} and Angle Resolved Photoemission Spectroscopy (ARPES) \cite{Zhou06} measurements suggest the existence of Dirac like carriers very close to the H point and massive carriers at the K point of the electronic band structure of bulk graphite. However, their contribution to transport using low-temperature and low-field magneto-transport measurements with advanced SdH oscillation analysis is still a matter of debate \cite{Schneider09, Igor06, Igor04, Mikitik06}. In the present study, the experimental accuracy and the limited number of SdH periods do not allow to confirm the presence of Dirac-like carriers in graphite at low field.

At higher magnetic field, the longitudinal resistance tends to saturate before developing an noticeable kink. This saturation has been explained in terms of freeze-out effect of the ionized impurity scattering centers in \cite{Iye85} whereas the formation of CDW accounts for the following resistance anomaly. The temperature dependence of the magnitude of the resistance jump as well as the following upturn are very similar to previous studies \cite{Yaguchi98}, although the smooth character of this anomaly is in contrast with the reported sharp sub-structures which are absent in our data. This effect is certainly related to disorder and/or voltage probe separation length which broadens the magneto-resistance fine details. Interestingly, the Hall resistance shows a sign change from negative to positive at magnetic field $B_R\approx 40$$\pm$1T, in the close vicinity of the magneto-resistance jump. However, in contrast to $B_c$, the Hall resistance sign inversion $B_R$ is temperature independent. Surprisingly, this effect continues to develop at temperature higher than 10K even when the CDW signatures in $R_{xx}$ have totally disappeared, suggesting a independent origin for these two effects. However, in the low temperature regime and very high magnetic field, the Hall resistance shows an upturn which is tentatively associated with the onset of re-entrance into the normal state. We note that the observed Hall resistance sign change is quite similar the one observed in other layered materials ($NbSe_{3}$, $(TMTSF)_{2}$ClO$_{4}$) in the quantum limit and has been suggested to result for quantum oscillations \cite {Uji05}. Another possible explanation lies in the electron-hole imbalance in high magnetic field as discussed later.

In an attempt to improve the details of the magneto-resistance anomaly and motivated by the apparent non-correlativity between the longitudinal and Hall resistances, a similar set of high field measurements have been performed with the same sample but with reduced dimensions between the electrodes (sample B). Similar results were observed and the experimental temperature range was extended to $300K$ (Fig.\ref{fig1}-c and d). The sign reversal of the Hall resistance is visible up to room temperature, however the magnetic field at which it takes place is temperature dependent for $T^*>20K$. Remarkably, $T^*$ acts as a threshold temperature above which quantum effects (SdH oscillations and plateau-like features) as well as CDW signatures all vanish.

At low temperature and high magnetic field, the positive Hall resistance above (B$_R$) field suggests that transport is dominated by hole-like quasi-particles. Earlier magneto-transport and theoretical calculations revealed that the Fermi surface of graphite consists of majority electrons as well as two hole pockets \cite{Williams65, John71}. The effective mass of electron and two types of holes were estimated to $m^{h}_{e}$$\approx$ 0.056 m$_o$ \cite {Dresselhaus02}, $m^{h}_{h}$$\approx$ 0.084 m$_o$ \cite {Dresselhaus02} and  $m^{l}_{h}$$\approx$ 0.003 m$_o$ \cite {Dillon77}, respectively. Recent measurements also suggest the existence of massless Dirac fermions (holes) near the H point of the Brillouin zone of graphite \cite{Orlita08, Igor06}. Classically, in such many carrier systems, the Hall sign conversion mainly depends on the respective magnitude of the electron/hole density $n_e$ and $n_h$ as well as mobility $\mu_e$ and $\mu_h$ ($\mu$=e$\tau$/$m^*$, where $e$ is the electron charge, $\tau$ is the scattering rate and $m^*$ is the effective mass). Thus, one may reasonably suppose that the hole density and/or mobility takes over the electronic-like carriers at sufficiently large magnetic field and low temperature, thus accounting for the temperature dependent behavior of $R_{xy} (B)$ and $B_R$ above $T^*$. Similar mechanism has been proposed for the Hall sign change in high magnetic field in a quasi-two-dimensional layered semimetal \cite{Everson84} and simple metals (Al, In) \cite{Achcroft69}. 

The band structure and Landau Level formation in a magnetic field applied perpendicular to the graphene planes have been first investigated by Slonczewski, Weiss and McClure (SWM model) \cite{Slonczewski58, McClure60, Nakao76} for pure bulk graphite. Recent studies suggest that it is very difficult to have defect-free graphite samples and the presence of impurities cannot be ruled out in experimental investigations \cite{Barzola08, Arndt09, Gonzalez07, Yonghua06}. However earlier magneto-transport measurements on natural graphite and HOPG are very reproducible and successfully confirm the band structure evolution of graphite at moderate magnetic field. In particular, the defect-free SWM model has been recently used to describe very detailed magneto-transport data in bulk graphite \cite{Schneider09}. Therefore, we may reasonably expect that defect-free theoretical models for graphite can account, on average, for the macroscopic bulk magneto-transport effects in a disordered and micro-structured graphite sample. The presence of impurities and lattice defects would essentially act as charge donors, providing charge carriers without significantly perturbing the basic electronic properties of bulk graphite. In strong magnetic field, the SWM model predicts that only a few Zeeman-split Landau sub-bands are populated, specifically $n=0^{\uparrow,\downarrow}$ (electron) and $n=-1^{\uparrow,\downarrow}$ (hole) bands (where $\uparrow,\downarrow$ denote the up and down spin, respectively) and no drastic magneto-effects should be expected for magnetic fields higher than 20T. The origin of magneto-resistance anomaly in graphite, experimentally observed at $B\approx 40 T$, cannot be accounted within the SWM model. Later, this model was refined by Takada et al. \cite{Takada98} by taking into account self-energy corrections and concluded that the $n = 0^\uparrow$ and $n = -1^\downarrow$ Landau subbands cross the Fermi level upwards and downwards respectively and almost simultaneously at $B=53 T$. Although this model has been proposed to explain the observed reentrance at $53 T$ \cite{Yaguchi98}, it fails to describe properly most of the experimental features. Indeed, the transition-like behavior of the high field magneto-resistance anomaly stimulated Yoshioka and Fukuyama \cite{Yoshioka81} to propose the formation of CDW in the electronic $n=0^\uparrow$ (electron) sub-band. Sugihara et al. improved the YF theory considering the screening effect on Coulomb interaction and concluded that the hole levels are responsible for the CDW formation \cite{Sugihara84}. This last scenario is consistently supported by the experiment reported in this letter, where the magneto-resistance anomaly takes place in a finite low temperature range (1.6K  to  10K), only when hole-type carriers dominate transport.

\begin{figure}[!ht]
\includegraphics[width=\columnwidth]{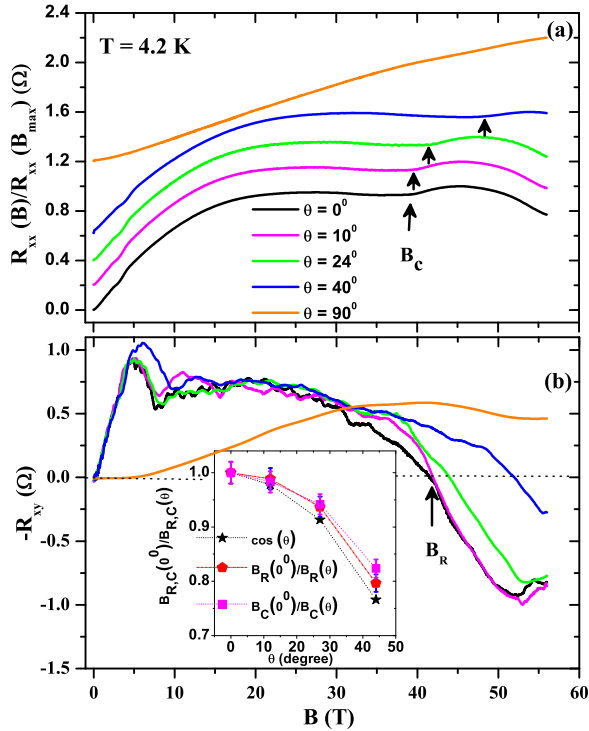}
\caption{\label{fig2} Angular dependence of the longitudinal (a) and the Hall resistance (b) at 4.2 K. In figure 2 (a), the R$_{xx}(B)$ resistance values have been normalized and successively shifted of 0.2 $\Omega$ for $\theta=0^o$ to $\theta=40^o$ and of 1.2 $\Omega$ for $\theta=90^o$ with respect to the reference curve at $\theta=0^o$ for clarity.}
\end{figure}

Figure \ref{fig2} shows angular dependent R$_{xx}(B)$ and R$_{xy}(B)$ measurements for a wide range of angles from $\theta=0^o$ to $\theta=90^o$.  In figure 2 (a), the R$_{xx}(B)$ resistance values have been normalized and successively shifted of 0.2 $\Omega$ form $\theta=0^o$ to $\theta=40^o$ and of 1.2 $\Omega$ for $\theta=90^o$ with respect to the reference curve at $\theta=0^o$ for clarity. It is noticed that the field at which CDW occurs (B$_c$) increases with increasing angle, at constant temperature (4.2 K). The Hall resistance shows a sign change from negative to positive at field (B$_R$) which also increases with increasing angle. When the magnetic field is parallel to the sample ($\theta=90^o$) R$_{xx}(B)$ does not show any field induced CDW feature while Hall resistance (R$_{xy}$) sign reversal is not observed upto the maximum field used in the present study. Theoretical calculations for the angular dependence of B$_c$ in graphite predicted that at constant temperature, B$_c$($\theta$) is approximately given by B$_c$(0)/B$_c$($\theta$) = cos ($\theta$), as only the magnetic field component along c-axis of graphite should contribute for the CDW instability \cite{Yoshioka81}. This is confirmed by the inset fig. 2 (b). It is pointed out that both B$_c$ and B$_R$ follow magnetic field component along the c- axis of graphite, within the experimental errors. When the field is applied parallel to the graphene planes ($\theta=90^o$), the Hall resistance is first zero for B$<$5T before increasing for higher magnetic field. This effect can be understood in terms of spin-polarized electron system and has been extensively discussed in the literature for the case of 2D electron gas \cite{Vitkalov01, Hwang06, Piot09}.

Recently, interlayer tunneling measurements on graphite have revealed the presence of a pseudo-gap in the quasi-particle density of states in presence of a magnetic field oriented along the c-axis \cite{Latyshev08}. Such a feature is shared by layered materials $NbSe_{3}$, $2H-TaSe_{2}$, $2H-Cu_{0.2}NbS_{2}$ etc.) \cite{Latyshev05, Evtushinsky08} and High Temperature SuperConductors (HTSC) \cite{Kondo09}. Actually, highly anisotropic layered materials ($NbSe_{3}$, $TaSe_{2}$ etc.) \cite{Coleman90, Everson84} and quasi-one-dimensional organic conductors ($(TMTSF)_{2}X$, where X is $PF_{6}, ClO_{4}$ etc.) \cite{Osada92, Graf04, Uji05} are known to exhibit CDW transition under high magnetic field. All these CDW bearing compounds materials exhibit a change of Hall coefficient sign soon after the CDW transition \cite{Coleman90, Everson84, Uji05}. Theoretical calculations demonstrate that a strong magnetic field  lead to the imperfect nesting of the Fermi surface and opens a gap in the electronic spectrum at the Fermi level \cite{Balseiro85} which may result the Hall sign reversal in highly anisotropic conductors \cite{Coleman90, Everson84,Evtushinsky08}. A similar sign change of the Hall coefficient has recently been discovered in HTSC \cite{LeBoeuf07}. These reported common features (occurrence of CDW, pseudogap and Hall sign reversal) point towards a generic property of highly anisotropic layered materials to which graphite belongs. 

To conclude, the present work reports on the magneto-resistance anomaly and Hall resistance sign reversal in HOPG under very high magnetic field. These experimental findings are consistent with the occurrence of CDW involving the hole sub-band in the electronic spectrum of graphite. The similarities with other layered materials exhibiting CDW formation, pseudo-gap and Hall resistance sign reversal are striking. In HOPG, the Hall resistance sign reversal sustains up to room temperature while the CDW feature disappears above $T\approx 10K$. Although these two phenomena take place at similar (but not exactly equal) magnetic field, the reported measurements suggest an independent origin. Contrary to other layered materials for which the Hall sign reversal has been theoretically studied, further investigations and dedicated theoretical calculations are required to understand the high magnetic field transport properties of HOPG.

This research is supported by EuromagNet II under E. U. contract 288043 (Proposal No. TSC 33-208) and ANR-JCJC-0034-01 research project. A. K. gratefully acknowledges to D. K. Avasthi, IUAC, New Delhi for fruitful discussions.

\end{document}